\documentclass[aps,pra,a4paper, amsfonts, amssymb, amsmath, reprint, showkeys, nofootinbib, twoside,onecolumn,superscriptaddress]{revtex4-2}
\usepackage[english]{babel}
\usepackage[utf8]{inputenc}
\usepackage[colorinlistoftodos, color=green!40, prependcaption]{todonotes}

\usepackage{longtable}
\usepackage{xspace} 
\usepackage[pdftex, pdftitle={Article}, pdfauthor={Author}]{hyperref} 
\usepackage{xr}

\usepackage{amsthm}
\usepackage{mathtools}
\usepackage{physics}
\usepackage{xcolor}
\usepackage{graphicx}
\usepackage[left=23mm,right=13mm,top=35mm,columnsep=15pt]{geometry} 
\usepackage{adjustbox}
\usepackage{placeins}
\usepackage[T1]{fontenc}
\usepackage{lipsum}
\usepackage{csquotes}
\usepackage{tabularx}
\usepackage{xspace}

\newcommand{\yearspan}{1986 and 2022\xspace}
\newcommand{\ncountries}{202\xspace}
\newcommand{\nproducts}{215\xspace}

\newcommand{\nsimu}{20\xspace}

\newcommand{\numReps}{20\xspace}
\newcommand{\negDemand}{$1.2\%$\xspace}

\makeatletter

\def\p@subsection{}

\makeatother

\begin{document}
\title{Diversification of global food trade partners increased inequalities in the exposure to shock risks}

\author{Ariadna Fosch}
    \email[Correspondence email address: ]{arifosch@gmail.com}
    \affiliation{Institute for Biocomputation and Physics of Complex Systems (BIFI), University of Zaragoza, Spain}
    \affiliation{Department of Theoretical Physics, University of Zaragoza, Spain}
    
\author{Alberto Aleta}
    \affiliation{Institute for Biocomputation and Physics of Complex Systems (BIFI), University of Zaragoza, Spain}
    \affiliation{Department of Theoretical Physics, University of Zaragoza, Spain}
\author{Roger Cremades}
    \affiliation{Sustainability Research Institute, School of Earth, Environment and Sustainability, University of Leeds, UK}
    
\author{Yamir Moreno}
    \affiliation{Institute for Biocomputation and Physics of Complex Systems (BIFI), University of Zaragoza, Spain}
    \affiliation{Department of Theoretical Physics, University of Zaragoza, Spain}
\begin{abstract}
Recent global food trade disruptions have evidenced how local shocks can cascade into global security threats. While the capacity of food systems to absorb spillovers depends heavily on its underlying trade networks, few studies quantify how their temporal evolution reshapes systemic vulnerability over time. Here, we evaluate how changes in global connectivity from 1986 to 2022 reshaped responses to production shocks. Using FAO data, we built yearly multiplex representations of the food trade system and quantified robustness through a stochastic shock-propagation model with dynamic export bans. We find that while increasing globalization intensified inter-dependencies and amplified cascades, robustness trends remain heterogeneous. Grain trade has become more decentralized and resilient to targeted shocks; conversely, Animal and Vegetable Fats exhibit growing centralization and fragility around key exporters like Indonesia and Malaysia. These structural transformations caused diverging shifts in systemic vulnerability, disproportionately threatening already vulnerable regions such as Africa and Southern Asia.
\end{abstract}

\keywords{food trade, production shocks, stochastic propagation model}
\maketitle

\section{Introduction}
In the last decades, the availability, stability, and access dimensions of food security have been critically challenged by several major disruptions in the food trade system~\cite{mcquinn2022quarterly, pingali2005food,nes2025economic}. For example, in 2022, the combined effects of severe droughts, the war in Ukraine, and the socioeconomic consequences of the COVID-19 pandemic led to a severe global food crisis~\cite{mcquinn2022quarterly,falkendal2021global}. The resulting food shortages and price inflation, reaching $70\%$ for cereals, led to a sharp increase in acute food insecurity, with the number of people at risk rising from 193 million in 2021 to nearly 258 million by 2022~\cite{nes2025economic}. Although prices began to stabilize during 2023, this episode illustrates both the magnitude and the speed at which stress can propagate through the interconnected global food trade network. 

The capacity of the food system to quickly absorb such disruptions~\cite{gribble2001robustness, bullock2017resilience}, i.e., its robustness, is not a static property; as it depends greatly on the structure of the trade networks. Shaped by economic development, globalization, and evolving geopolitics, global food trade networks have grown in scale and complexity~\cite{carr2016commodities}. However, it remains unclear whether these changes amplified or created new pathways for food system shocks to spread and cascade across countries~\cite{sartori2015connected}. Characterizing how the food trade network’s structure has evolved and how this has impacted the propagation of production shortages could provide crucial insights into the resilience of the food system against future cascading shocks.

Previous research has typically addressed the robustness of the food trade system from two complementary perspectives. Temporal evolution analyses examined how the network’s structure changes over time, characterizing shifts in connectivity~\cite{yin2024temporal, torreggiani2018identifying,hu2024evolution}, community formation~\cite{wang2023trade,torreggiani2018identifying}, efficiency~\cite{karakoc2021complex}, and topological robustness~\cite{karakoc2021complex, jafari2024multiple,ma2023robustness}. Shock propagation studies, in turn, simulated how local shortages spread across trade networks during crises such as the COVID-19 pandemic~\cite{swinnen2020covid}, extreme climatic events~\cite{puma2015assessing}, or wars~\cite{laber2023shock,bertassello2023access,kuhla2024international}. Yet, a complete assessment of robustness would require integrating the exploration of both structural and shock propagation dimensions.

In this study, we combine both perspectives to analyze the evolution of international food trade networks between \yearspan and assess how structural transformations over time have affected the propagation of production shocks. Unlike previous studies, which primarily focus on staple crops like wheat~\cite{ma2023robustness,puma2015assessing,yin2024temporal} or rice~\cite{puma2015assessing}, we explore 12 distinct food categories with heterogeneous trends of robustness. This is possible thanks to the innovative use of a multiplex network framework, encoding food categories as separate layers sharing a set of nodes that represent countries. 

We first analyzed the temporal evolution of the food trade multiplex, revealing the most relevant structural transformations that affected food trade systems between 1986 and 2022. We then introduce a novel stochastic model to simulate the propagation of production shortages across the food-trade multiplex. This model simulates how countries facing product shortages react by dynamically restricting their exports, thus limiting the availability of goods for dependent partners, and creating new shortages that ripple through the network. Who is affected by such restrictions is defined through a stochastic propagation rule that proxies competition between importers. These dynamics reflect competition under export restrictions based on taxes, quotas, or bans typically applied by major exporters to stabilize their domestic markets~\cite{akter2022effects}. For example, during the 2008–2011 food crisis, prices spiked between 13\%–40\%. This caused countries like Vietnam, India, Egypt, China, and Cambodia to apply rice export bans to protect their own domestic consumption~\cite{headey2023food,nes2025economic,ma2023robustness}. India also relied on export bans in July 2023, when it applied a global export ban for all non-basmati rice to prevent local price inflation~\cite{nes2025economic,baum2024adaptive}. This restriction is estimated to have led to losses of about $315$ million dollars for global rice consumers~\cite{nes2025extreme}. 

Finally, we applied this model to estimate the spillover cascades derived from simulated production shortages between 1986 and 2022. For each year and food category, we simulated the spillover cascades derived from single-country shocks independently affecting each global producer. Then, we compared the size of the generated spillover cascades over time. Our approach not only quantifies how changes in trade structure affect the spread of crises, but also reveals shifts in the most vulnerable geographical regions for each food category. This provided a dynamic view of the resilience of the global food trade network over time that could be of special interest for policymakers aiming for global food security. 

\section{Results}\label{results}

\begin{figure}[ht]
    \centering
    \includegraphics[width=0.85\linewidth]{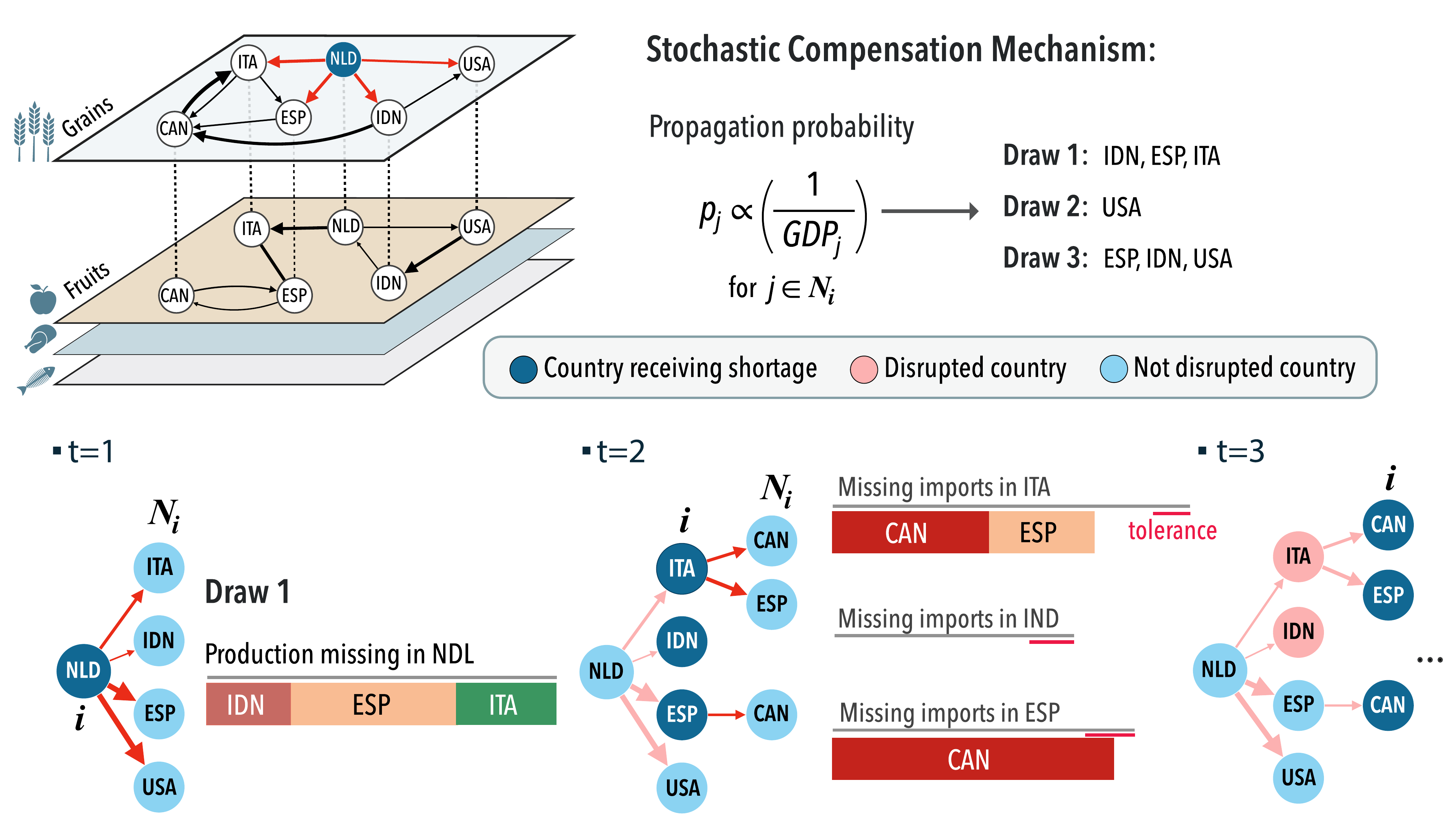}
    \caption{Schematic illustration of the food trade multiplex and the stochastic shock–propagation model. For each year, the food trade system is represented as a multiplex network with 12 layers, each corresponding to a food category, where nodes are countries and directed links are trade flows. Shock propagation is simulated independently within each layer. A production shortfall is introduced in one country (e.g., the Netherlands, NDL), which aims to compensate for it by reducing exports to a subset of its partners ($N_i$) selected through a stochastic mechanism weighted by partner GDP. In the example, reducing exports to Indonesia (IDN), Spain (ESP), and Italy (ITA) is sufficient for NDL to fully compensate its deficit at $t=1$, allowing the United States of America (USA) to maintain its regular supplies. At $t=2$, the countries affected by export reductions experience supply shortages, and thus, they will apply their own compensation mechanisms. Even after reducing all exports, both Italy and Indonesia will fail to cover their demand. Thus, we consider them to be ``Disrupted'' by the shock. Contrarily, applying its own export reductions allows Spain to cover most of the missing imports, bringing the shortage into a ``tolerable'' range and preventing the disruption of the country. Even if Spain is not considered disrupted by the shock, it had to implement export reductions, thus spillover effects will still propagate to Canada (CAN) at $t=3$. The shock propagates through the network until no more countries are disrupted by it. }
    \label{fig:scheme_model}
\end{figure}

Using bilateral trade flow matrices from the FAO, we built yearly multiplex representations of the trade of food products between \yearspan (see Sec.~\ref{meth:multiplex} for more details). Through the multiplex framework, we can represent multiple networks connecting the same set of nodes (countries) in a single structure with separate layers (see Fig.~\ref{fig:scheme_model} for a schematic representation). The resulting network is directed: outgoing links capture a country's exports, which implicitly defines the incoming links as the corresponding imports for the receiving country. The multiplex consists of 12 layers, each reflecting these directed trade relationships for one macro Food-Ex category (e.g., Grain products, Meat products, Animal and Vegetable Fats)~\cite{fao1993codex}. Finally, we built two weighted versions of this multiplex: one weighting the links by economic value, used for topological analyses, and another weighting the links by trade volume, used for shock simulations.

We then used this yearly networked representations of international food trade to study: (1) how the topological changes in the multiplex (e.g. increase/decrease in trade volume, changes in trade partners, changes in dominance) have evolved over time; (2) how these topological changes have altered the propagation of production shortages through the multiplex; and (3), whether these topological changes have altered the geographical distribution of vulnerability.

\subsection{Trade expanded, dependence on imports grew}\label{sec:temporal_evo}
\begin{figure}
    \centering
    \includegraphics[width=\linewidth]{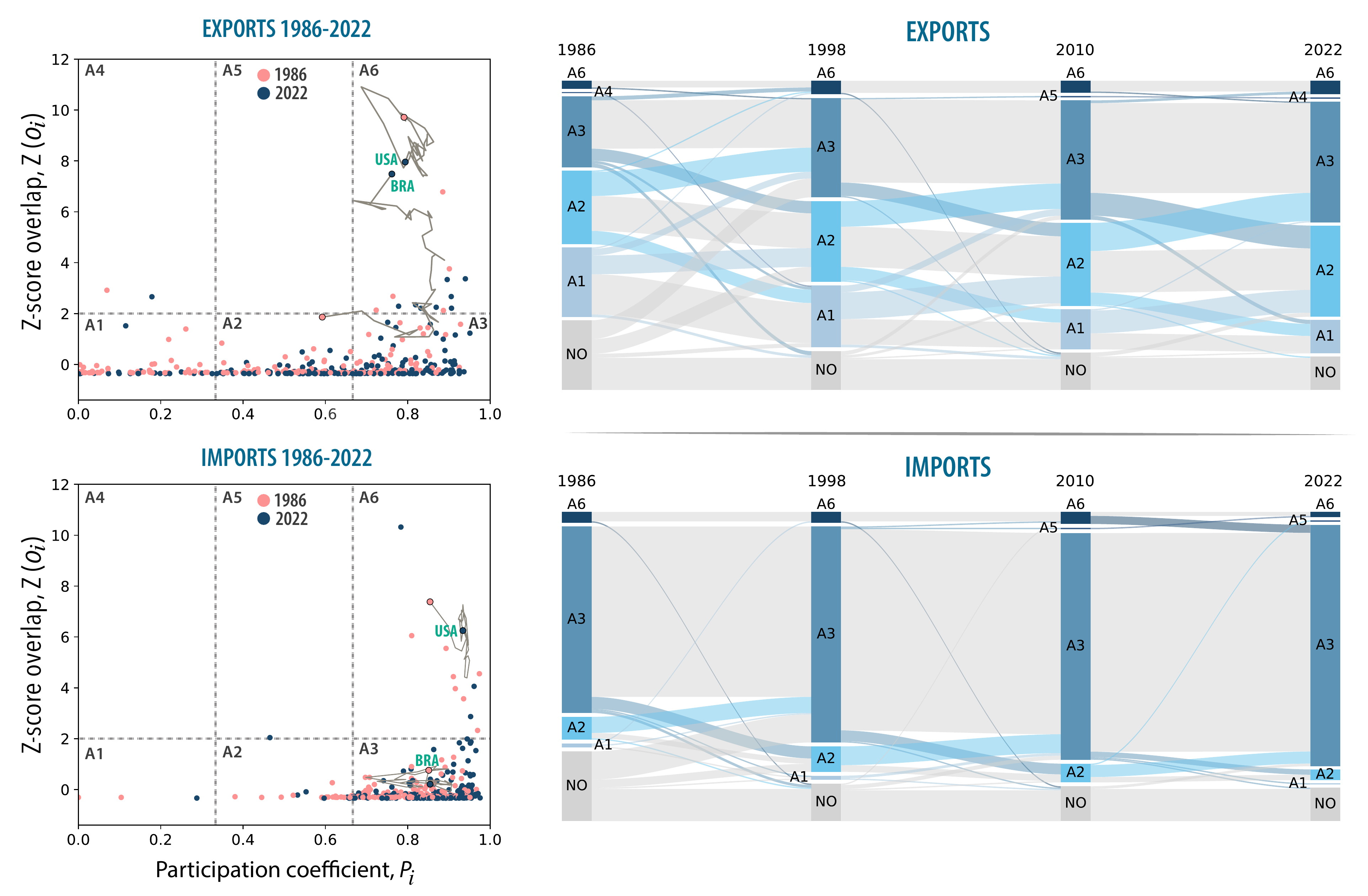} %updated2023
    \caption{Temporal evolution of the food trade multiplex between \yearspan. The left panels show the position of countries in the Multiplex participation coefficient ($P_i$) and weighted overlapping degree ($Z(o_i)$) plane, both for exports (upper) and imports (lower). These two panels show the position of all countries for the first and last year studied, 1986 and 2022 (pink and blue, respectively), and the complete developmental trajectories for two example countries, Brazil (BRA) and the United States of America (USA). The $Z(o_i)-P_i$ plane is divided into six sectors, labeled following the categories defined in Sec.~\ref{sec:plane}:  A1 ``Localized nodes'', A2 ``Mixed nodes'', A3 ``Multi-product nodes'', A4 nodes are considered ``Localized hubs'', A5 ``Mixed hubs'', A6 ``Multi-product hubs'', NO ``In-existing countries in that year'' (i.e., countries not yet been created /countries that disappeared). Meanwhile, the right panels present the flow of countries across the 6 sectors of the $Z(o_i)-P_i$ plane at four different years between \yearspan. Note that this analysis has been performed using a multiplex where edges are weighted by the monetary value of the transaction (in 1000 USD).}
    \label{fig:exports}
\end{figure}

Globalization has profoundly reshaped the global food trade network, expanding both the number of participating countries and the diversity of products exchanged. These structural transformations were captured through the temporal evolution of two multilayer network metrics: the Z-score of the weighted overlapping degree $Z(o_i)$ and the multiplex participation coefficient $P_i$~\cite{battiston2014structural}. The former quantifies the total trade volume of country $i$ across all food categories (exports or imports), while the latter indicates whether this trade is concentrated in a few categories ($P_i=0$) or evenly distributed across many ($P_i=1$). Together, these metrics define the $Z(o_i)$–$P_i$ plane (see Fig.~\ref{fig:exports}, left panels), which we divide into 6 sectors (A1-A6) to classify countries according to their level of trade dominance and product diversification (see Sec.~\ref{sec:plane} for more details on the classification). Tracking how the number of countries in each sector changes over time allows us to visualize the topological transitions in countries’ trade dominance (Fig.~\ref{fig:exports}, right panels).

Between \yearspan, we observe a clear shift toward a more interconnected and diversified system (see Fig.~\ref{fig:exports}, upper panels). The number of highly specialized exporters markedly declined, as many countries broadened their export portfolios and strengthened their roles within the global network. This diversification is particularly evident among intermediate exporters (A2-A3 sectors), whose numbers have increased substantially at the expense of specialized exporters (A1). Although most major multi-product exporters (A6), such as France, the Netherlands, and the United States, have retained their central roles (see Supplementary Fig.~S1), new key players like Brazil and Germany emerged during the 1990s. This broader participation has gradually diluted the dominance of traditional hubs, evidencing the shift toward a more decentralized and interconnected global export system.

On the import side, while most nations already imported a variety of products in the 1980s, their reliance on imports to meet domestic demand has nearly doubled, rising from 23\% to 40\% by 2022 (see Supplementary Fig.~S2). Most countries now occupy intermediate positions in the network (A3), importing moderate volumes from multiple food categories, while highly specialized importers (A1 countries) have become increasingly rare. Interestingly, the number of large-scale importers (A6) remained stable for decades, though it declined slightly following the COVID-19 pandemic (see bottom-right panel of Fig.~\ref{fig:exports} and Supplementary Fig.~S1). Together, these trends indicate that globalization has driven the food trade system toward greater diversification and interdependence.

We further support our results by exploring the temporal evolution of the Inverse Participation Ratio (IPR)~\cite{goltsev2012localization}, see Supplementary Sec.~1.3. We observe a decline in the IPR over time for both types of trade, corroborating that exports have become more diversified and more countries are importing from each other.

\subsection{Production shortages in 2022 induce larger spillover cascades}\label{sec:cascade}

\begin{figure}[ht!]
    \centering
    \includegraphics[width=0.9\linewidth]{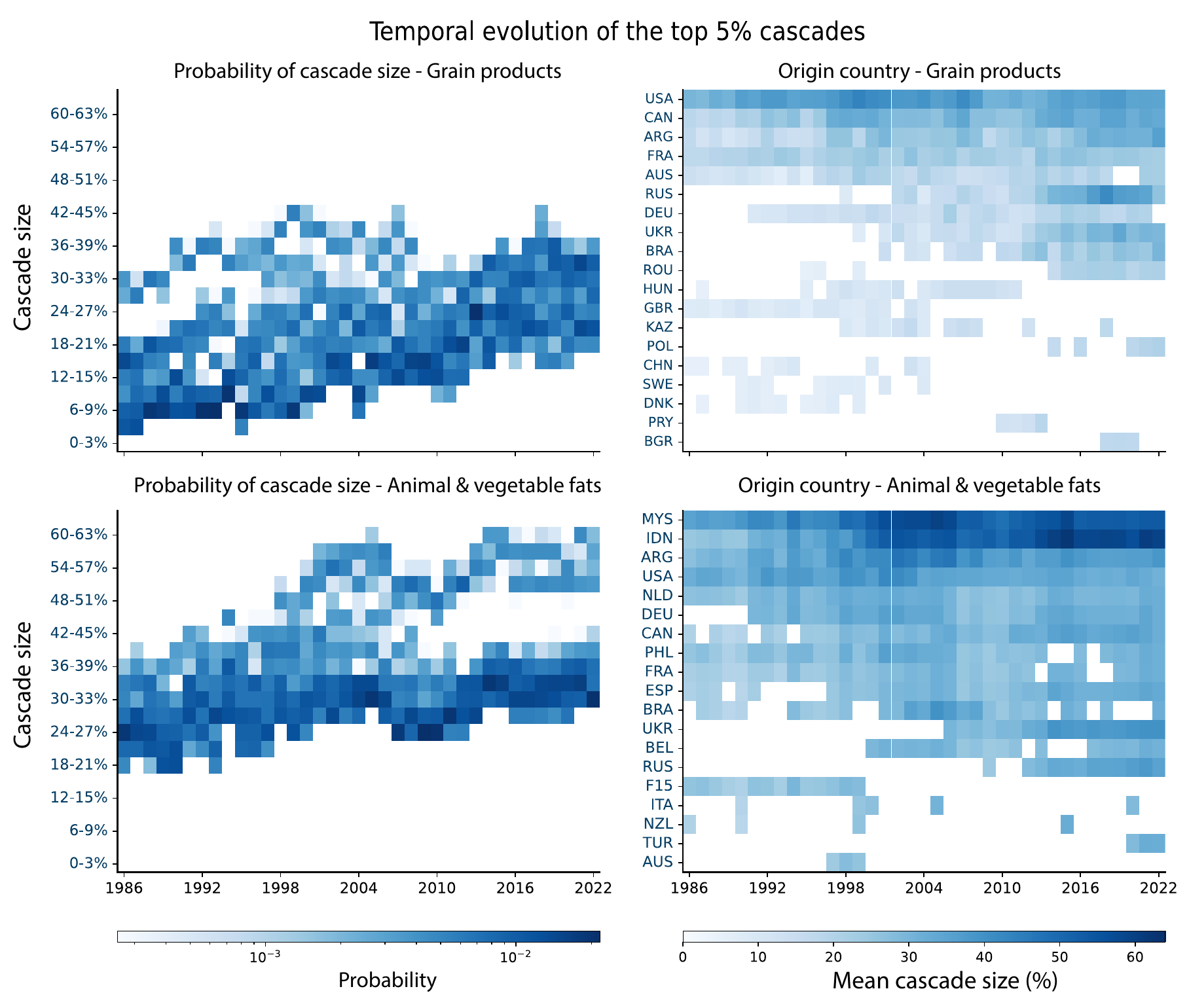} 
    \caption{Temporal evolution of the largest cascades (top 5\%) of each year between 1986 and 2022 for the Grain products and Animal and Vegetable Fats layers (upper and lower panels, respectively). Results are based on \numReps stochastic simulations of single-country shocks initiated in each producer country for each year and food category. The left panels show the temporal evolution of the cascade size. Color indicates the probability of having a cascade with a specific size. Meanwhile, the right panels show which countries are triggering the top 5\% cascades in each year. In this case, color indicates the average cascade size of the top cascades initiated by that country. For visualization purposes, the right panels have been limited to the top 19 countries (complete panels are available in Supplementary Fig.~S5).}
    \label{fig:cascade_size}
\end{figure}

With the food trade multiplex shifting toward a more densely interconnected structure, it is natural to also observe changes in the size of the spillover cascades derived from shocks. We quantified these effects by simulating the same type of shock across different years and evaluating cascade size changes. More specifically, for each year and food category, we used our stochastic propagation model to simulate the spillover cascades resulting from a 100\% production loss in each global producer, repeating each simulation \nsimu times to account for stochastic variability. We then measured the cascade size as the percentage of countries that, at any time-point of the simulation, fail to completely cover their shortage through export bans, and are left with a remaining deficit larger than a pre-defined tolerance threshold $b$. This threshold, set at 1\% of annual demand ($b=0.01$), reflects the ability of countries to buffer supply shocks (e.g., through stocks or adaptive measures). A spillover cascade of 50\%, for instance, indicates that half of the countries experienced shortages exceeding 1\% of their demand before the shock was fully absorbed. A detailed description of the shock propagation model and the cascade size calculation is provided in Sec.~\ref{sec:model}-\ref{meth:casc_size}.

The increase in import dependence and the expansion of trade during the 1990s translated into a general increase in the size of spillover cascades triggered by shocks. For Grain products, looking at the largest cascades in the system (top 5\% of the distribution), the minimum cascade size grew steadily over time, expanding from affecting just 4\% of the network in the 1980s to 20\% by 2022 (Fig.~\ref{fig:cascade_size}, upper-left panel). A similar trend is observed among the top cascades for the Animal and Vegetable Fats layer (Fig.~\ref{fig:cascade_size}, lower-left panel). Here, the minimum cascade size nearly doubled during the 1990s - rising from 18\% to 27\% - and stabilized thereafter. While both food categories exhibit an overall growth in cascade size, the shape of their size distributions evolved in opposite directions as a consequence of distinct adaptive transformations in the structure of their trade networks.

In the case of Grain products, the system underwent a decentralization phenomenon. In 1986, global grain trade was heavily dependent on a few dominant suppliers —most notably the United States (Fig.~\ref{fig:cascade_size}, upper-right panel). As a result, shocks originating in the U.S. triggered cascades affecting $30-33\%$ of countries, while disruptions in other major producers had smaller impacts, affecting less than $21\%$ of countries. As trade expanded and smaller suppliers gained dominance, these disparities diminished, and the previously bimodal cascade size distribution became unimodal (Fig.~\ref{fig:cascade_size}, upper-left panel). This transition reflects an evolution towards a more decentralized grain trade market, one where no single country is unique in triggering catastrophic cascades.

Contrarily, the Animal and Vegetable Fats layer underwent a centralization phenomenon. Cascade sizes grew uniformly during the 1990s, but after 2004 the cascade size distribution bifurcated: shocks originating in Indonesia and Malaysia generated extremely large cascades affecting $50-63\%$ of countries, while shocks in other top producers produced more moderate impacts, $33-45\%$ (Fig.~\ref{fig:cascade_size}, lower panels). This bifurcation mirrors the rise of the palm oil industry, which came to dominate the global vegetable oil market but remains geographically concentrated in just a few tropical countries, mainly Indonesia and Malaysia~\cite{meijaard2024exploring}. Such spatial concentration heightens systemic vulnerability (e.g. to a new pest or a geopolitical risk), exposing the trade of oils to severe disruptions when key countries are affected (see Supplementary Fig.~S6).

These two examples illustrate the most common adaptive transformations of the trade networks (centralization and decentralization) observed across the 12 food categories explored (see Supplementary Fig.~S7 for all categories). Most categories followed a similar trajectory to Grains, evolving towards a more decentralized trading system where all the countries originating the largest spillover cascades have similar impacts. Meanwhile, the trade for Sugar and confections became increasingly centralized, with Brazil emerging as a dominant hub capable of triggering catastrophic cascades and other countries generating more modest perturbations. Finally, locally produced and consumed goods, such as Eggs, remain largely insulated from global shifts, as they do not present relevant topological changes that modify shock propagation. 

\subsection{Vulnerability shifts following topological changes}\label{sec:vulnerability}
\begin{figure}[ht]
    \centering
    \includegraphics[width=0.9\linewidth]{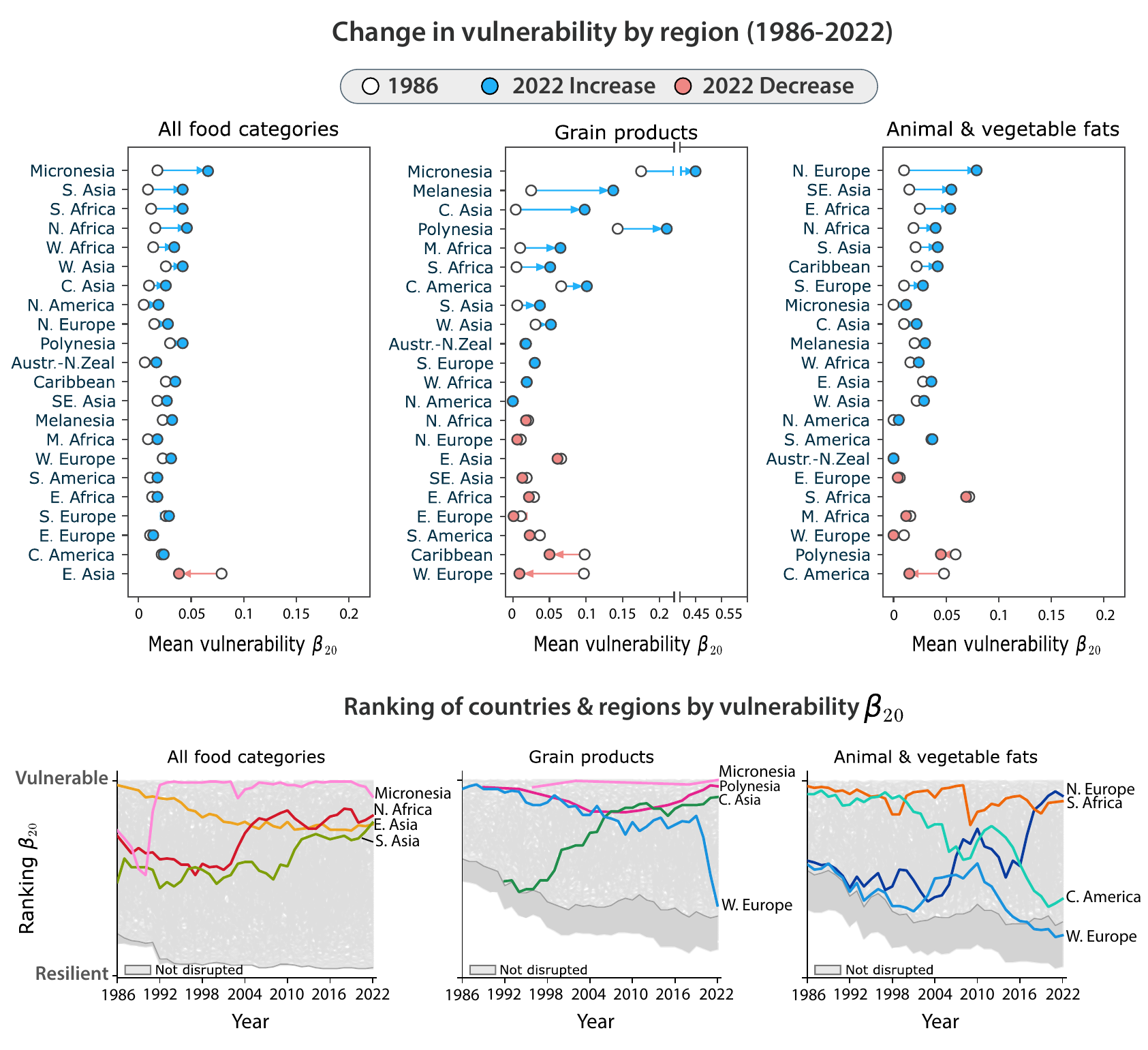} %updated_w4
    \caption{Analysis of the temporal evolution of vulnerability ($\beta_{20}$) for all UN geographical sub-regions. Upper panels show the change in average $\beta_{20}$ in each UN subregion between the first (white) and last (colored) year of data available. For most cases, this corresponds to 1986 and 2022. When trade data was not available, we considered the first year with data: Central Asia (1992), Micronesia (1996), and Polynesia (1989) for Grains, Central Asia (1993) and Micronesia (1989) for Fats; and  Central Asia (1992) for all food categories. Arrows are colored based on the direction of the vulnerability change, blue for an increase and pink for a decrease. The lower panels show the temporal evolution of the vulnerability ranking between 1986 and 2022. Gray lines reflect the individual trends of each country, while we have colored the average response in some sub-regions of interest. The darker gray area reflects the countries that are not disrupted by any spillover cascade, thus have $\beta_{20}=0$. In both analyses, geographical sub-regions follow the definitions from the United Nations (UN), see Supplementary Sec.~3.2 for more details.}
    \label{fig:ranking_vulnerability}
\end{figure}
Our results suggest that the evolution of the food trade multiplex has amplified both the potential scale and spread of spillover cascades from production shocks. Rising import dependence has increased the overall cascade size, while structural changes have shaped how these shocks propagate: centralization makes the system highly vulnerable to disruptions in a few key exporters, whereas decentralization allows a larger number of countries to trigger cascades. This raises the question of whether these transformations in the network structure have also changed which countries are the most vulnerable against spillover cascades. 

In this work, we define vulnerability as the limited capacity of a country to buffer supply disruptions. For example, India stores 40–50 million tons of rice and wheat under its National Food Security Act~\cite{USDA_IN2025_0023}. These stocks should help it reduce the impact of spillover cascades, increasing India's robustness against supply shortages. With this in mind, we quantify vulnerability with a measure of how much a country should increase its reserves to survive $X\%$ of the shocks that otherwise would disrupt them; we call this measure $\beta_{X}$. Intuitively, countries with a $\beta_{X}\approx0.01$ are less vulnerable, as they would only need to minimally increase the assumed tolerance $b$ to reduce by $X\%$ the non-compensated shortages. Meanwhile, countries with $\beta_{X}\approx1$ are more vulnerable, as they would need to stockpile unfeasible amounts of product to reduce their exposure by $X\%$. In this study, we focus on $\beta_{20}$ as a realistic benchmark for national preparedness, and we provide similar results for $\beta_{50}$ in Supplementary Sec.~2.4. 

To better characterize the temporal evolution of vulnerability, we estimated $\beta_{20}$ for all countries, years, and food categories, and explored its time series. We filtered out local oscillations by smoothing the series with a 4-year rolling window and then aggregated individual country estimates into geographical sub-regions~\cite{moving_average}. Overall, we see that vulnerability increased between 1986 and 2022 in nearly all regions except for Eastern Asia (Fig.~\ref{fig:ranking_vulnerability}, upper-left panel). This growth, however, has not been homogeneous. Regions such as Micronesia, Southern Asia, and much of Africa present a marked vulnerability increase, whereas for most of Europe and Central and South America, it only grew marginally. These inequalities are particularly relevant because they disproportionately affect critical regions for global food security. Further disaggregating these results by food category provides more nuances on these vulnerability shifts, revealing product-specific trends that may be of interest for policymakers (see upper-middle and upper-right of Fig.~\ref{fig:ranking_vulnerability} and Supplementary Fig.~S8).

Through our temporal analysis of the vulnerability ranking, we also could reveal when these vulnerability shifts occurred and how they relate to the topological changes in the multiplex structure (Fig.~\ref{fig:ranking_vulnerability}, lower panels). For example, finding Polynesia, Micronesia, or Melanesia at the high end of most vulnerability rankings is not a surprise. Isolated land-scarce regions are naturally more vulnerable to shortages, due to their limited productive capacity and high import reliance~\cite{fanzo2024global} (see Supplementary Sec.~2.6 for more details on the relationship between import dependence and vulnerability). 

Other regions, however, experienced vulnerability shifts that permanently altered their positions in the vulnerability ranking. We associate these long-term changes with transitions in the network structure that systemically conditioned the propagation of production shortages. For example, Central Asia rose significantly in the vulnerability ranking during the decentralization of the grain trade network in the 90s and early 2000s. We also observed similar long-term vulnerability shifts in the Animal and Vegetable Fats case, where we observed a constant growth in vulnerability for Northern Europe and a steady decline in Central America. These shifts reflect, respectively, the growth in imported oils of European countries~\cite{gunstone2011production}, and the production increases in Central America~\cite{furumo2017characterizing}. 

Besides long-term vulnerability shifts, we also observed transient perturbations in the vulnerability ranking. These are fast shifts in the vulnerability ranking that return to previous levels afterwards. We believe that transient shifts reflect the response of the system to specific events that perturbed trade relationships, but do not result from stable topological changes in the structure of the multiplex. For example, the reduction of vulnerability in Western Europe during the 2019-2022 period possibly relates to a global decrease on grain trade during the Covid-19 pandemic and the subsequent Russia-Ukraine war~\cite{USDA_EU2022}. We expect this to be a transitory vulnerability shift, but time will tell whether the impacts of such crises are enough to cause persistent vulnerability transitions. 

In short, between 1986 and 2022 there has been a global increase in vulnerability to production shortages. This growth has been heterogeneous and can be linked both to stable structural transitions observed in the trade multiplex and other transient effects. Our analysis also uncovered growing inequalities in the exposure to shock risks. Critical regions for global food security, such as Micronesia, Southern Asia, and most of Africa, presented the highest increases in vulnerability, which would significantly amplify their risk in the event of a global production shortfall.

\section{Discussion} 
Our novel insights on the temporal evolution of global food trade networks reveal heterogeneous centralization–decentralization patterns that altered shock propagation and led to a systemic increase in vulnerability. This increase, however, has been unevenly distributed across regions. Southern Asia and most of the African continent have disproportionally increased their vulnerability, while Eastern Europe, Southern Europe, and Central America only increased it marginally. This uneven redistribution of vulnerability is particularly concerning, as it amplifies pre-existing inequalities. Here, inequalities in exposure refer to an increasingly uneven distribution of shock vulnerability across countries with markedly different capacities to absorb supply disruptions, rather than to differences in trade volumes or economic performance per se. For regions with limited food accessibility, like Micronesia or Africa, which benefit from the stabilizing effects of international trade~\cite{trade2025economic, puma2015assessing,resilience_report}, this comes at the expense of an increasing vulnerability to spillover cascades, which would dramatically worsen their situation if a shock were to manifest~\cite{mary2019hungry,fanzo2024global,meng2024larger,puma2015assessing,torreggiani2018identifying, mary2019hungry}. Mitigating this added risk is challenging from a policymaking perspective, since often countries do not have enough infrastructure to maintain the stocks required to protect themselves from food shortages. However, food storage has proven to be an effective intervention. For example, India includes the stocking of grain reserves in their national food security strategy. These stocks cover around 14\% of the grains demand~\cite{USDA_IN2025_0023}, which proves to be aligned with our vulnerability estimates. The vulnerability of Southern Asia has increased between \yearspan, but a buffer of grains covering 14\% of the national demand would protect India from more than 50\% of the shocks that it is currently vulnerable to ($\beta_{20}=0.037, \beta_{50}=0.046$ for Grain products). Knowing the country-level exposure to global systemic vulnerabilities is crucial for building adaptive capacity to systemic risks, which might originate from climate extremes, geopolitical tensions, market volatility, or other crises. The use of multilayer networks remains relatively limited in food trade research~\cite{baum2024adaptive,torreggiani2018identifying}, yet our results show that neglecting product heterogeneities can mask structurally different vulnerability regimes and lead to misleading conclusions about systemic resilience. 

While our framework captures key structural and dynamical features of global food trade, several limitations should be acknowledged. First, the shock-propagation model focuses on short-term crisis responses, assuming that countries react to production shortfalls primarily through export restrictions, without allowing for the formation of new trade links, endogenous demand adjustments, or longer-term production responses. This design choice reflects the timescale at which many food crises unfold, but it implies that our simulations should be interpreted as first-order stress tests of systemic exposure rather than forecasts of realized shortages. Second, food products are aggregated into broad FoodEx2 categories to ensure data completeness and policy relevance across decades; although this may obscure heterogeneity within categories, our results indicate that for actively traded products, finer disaggregation would likely amplify, rather than weaken, the centralization patterns we identify. Third, limited and uneven reporting of national food stocks prevents an explicit modeling of dynamic stock depletion, leading us to represent buffering capacity through an abstract tolerance parameter. Despite these limitations, the relative comparisons across time, food categories, and regions remain robust, allowing us to isolate how structural changes in trade networks have reshaped the geography of vulnerability to food supply shocks.

\section{Methodology}\label{sec:methods}
\subsection{Data acquisition and preprocessing}\label{data}
We relied on trade and production data provided by the Food and Agriculture Organization of the United Nations (FAO)~\cite{FAO_trade}. This takes the form of normalized annual bilateral trade flow matrices and production records for 220 countries and 559 crop and livestock products between the years 1986 and 2023. 

The production dataset, on the other hand, is only reported in weight (tons) of product. Moreover, it is only available for a smaller subset of the traded products. Thus, for this study, we only considered the subset of \nproducts products that are available in both datasets for all years between \yearspan. 

Bilateral trade-flow matrices combine import and export records. Thus, it is common to observe the same trade flow reported twice, by the seller and the buyer. We observe double reporting in 60\% of the records, with a mean (absolute) relative difference in the magnitude reported of $61\%$, across all goods, countries, and years. We addressed it by considering the transaction report from the most reliable country. Reporting reliability is measured per commodity and year, following a data-based approach adapted from Gehlhar et al.~\cite{gehlhar1996reconciling}. 

Other preprocessing steps involved: 
\begin{itemize}
    \item Removing all records with the same origin and destination country, as they refer to the national trade of ``special goods''.
    \item Removing livestock products reported in ``heads'' or ``Number'' units (approx 1\% of the data). 
    \item Removing the Rice paddy category, as it is an aggregated class combining the records of other rice products. 
    \item Representing all trade flows as export relationships.
    \item Removing by-products from food production that are not suitable for human consumption (e.g., hay or leather). 
\end{itemize}

In the end, we obtain a reconciled trade matrix where all trade-flows reflect the total exports of a certain product between two countries for a specific year, both in terms of volume (tons) and economic value (1000 USD). 

\subsubsection*{Gross Domestic product}
We extracted yearly Gross Domestic Product (GDP) data in US dollars from the FAO’s Macro Indicators database~\cite{FAO_gdp}. The database's content is based on the countries' official National Accounts data reported through the annual National Accounts Questionnaire~\cite{FAO_gdp}. This database provides a complete time series for 194 out of the \ncountries countries available in the FAO trade data. It also has partial time series for 7 more countries, which were reconstructed through linear imputation, see Supplementary Sec.~3.1. We also estimated the GDP for "Belgium-Luxembourg" (ISO Code 3: F15) by adding up the GDPs of Belgium and Luxembourg for the years 1986-1999. 

In the end, we work with a subset of \nproducts products and \ncountries countries for which we have complete trade, production, and GDP records for all years between \yearspan. The complete list of countries and products considered can be found in Supplementary Sec.~3.2-3.3.

\subsubsection*{Food group aggregation}\label{sec:foodex}
We aggregated the individual food products from the FAO databases into more representative food categories~\cite{marchand2016reserves} using the FoodEx2 Exposure hierarchy database. FoodEx2 is a food classification and description system developed by the European Food Safety Authority (EFSA) for dietary exposure calculations~\cite{european2015food}. This classification system provides 7 different hierarchical aggregation levels, allowing us to categorize individual food products into groups and subgroups with increasing granularity. We manually mapped each product from the FAOSTAT database to one of the classes in the largest hierarchical level of FoodEx2 (L1). Then, for every pair of trading countries, we computed the total trade value within each food category by summing the magnitude of all transactions in that category. Production data was aggregated similarly. Our dataset contains samples for 12 out of the 20 classes of the L1-hierarchy, distributed as shown in Supplementary Tab.~S2. 

\subsection{Food-trade multiplex}\label{meth:multiplex}
Multiplex networks $M = (\mathcal{N},\mathcal{E},\mathcal{L})$ are an extension of the network formalism that allow us to represent $L$ networks connecting the same set of nodes $\mathcal{N}= \{n_i: i \in \{1, . . . , N\}\}$ as different layers $\mathcal{L} =\{l:\{1,...L\}\}$ of a single structure~\cite{cozzo2018multiplex,aleta2019multilayer}. In the multiplex network formalism, the set of edges $\mathcal{E}=\{(n_i^ln_j^{l'}): i,j\in{\mathcal{N}}\wedge  l,l^{'}\in\mathcal{L}\}$ can be separated into two types, intra-layer edges, reflecting the connections between nodes in the same layer ($l=l'$) and inter-layer edges, which are the coupling links between layers ($l\neq l'$). This allows us to study the influence of nodes inside each layer and their relevance in the system as a whole simultaneously. 

For our analysis, we used the pre-processed trade data from FAO to build a yearly multiplex representation of the food-trade system. Each version of the food trade multiplex $F^y=(\mathcal{N}^y,\mathcal{E}^y,\mathcal{L})$ contains the same set of layers $L=12$, reflecting the 12 food categories from the FoodEx2 classification~(see Sec.~\ref{sec:foodex}). However, the set of nodes $\mathcal{N}^y$ and edges $\mathcal{E}^y$ varies to reflect the set of countries and their trade relationships in year $y$. Each layer of the multiplex is a directed and weighted trade network $W^y_f$, where the nodes represent the countries existing in year $y$, and edges are the export relationships for that year in the food category $f$, weighted by transaction magnitude. Unweighted inter-layer links were also added to connect nodes that represent the same country in different layers. 

For each year, we created two versions of the food-trade multiplex, one weighting intra-layer edges by volume traded (tons) and another one weighting them by economic value (1000 USD). The economic-weighted multiplex was used for the topological analyses in Sec.~\ref{sec:temporal_evo} while the volume-weighed one was used for the shock propagation simulations (Sec.~\ref{sec:cascade} and~\ref{sec:vulnerability}).

Note that, in the multiplex formalism, the set of nodes $\mathcal{N}^y$ (countries) is constant across layers in the same year. Therefore, countries not participating in the trade for specific food categories are represented as isolated nodes in the corresponding layer. However, when comparing multiplexes of different years, the number of countries may differ. This reflects the creation/destruction of countries resulting from changes in the global political landscape. For simplicity, we consider each country as an independent entity. Thus, even if South Sudan split from Sudan in 2011, we treat them as independent nodes of the multiplex: ``Former Sudan'' (1986-2010), ``Sudan'' (2011-2022), and ``South Sudan'' (2011-2022).

\subsection{Plane $Z(o_i)-P_i$}\label{sec:plane}
In a directed multiplex, the weighted overlapping degree ($o_i$) is defined as the sum of the weight of all edges exiting ($o_{{\text{out}}_i} = \sum_{i}w_{i,j}^l$) or entering ($o_{\text{in}_i} = \sum_{j}w_{i,j}^l$) a node $i$ across all layers~\cite{battiston2014structural}. Therefore, it provides a measure of the relevance of a country as a global exporter ($o_{\text{out}}$) or importer ($o_{\text{in}}$) in all food products. Since the weighted overlap is sensitive to the total magnitude traded, we use its Z-score to compare the multiplex in different years. This is defined as 

\begin{equation}
    Z(o_i)= \frac{o_i-\langle o_i\rangle}{\sigma_{o_i}}.
\end{equation}
where $\langle o_i\rangle$ and $\sigma_o$ are respectively the mean and standard deviation of $o_{i}$ . 

On the other hand, the multiplex participation coefficient ($P$) is defined as 
\begin{equation}
    P_i= \frac{L}{L-1} \cdot \left( 1- \sum_{l=1}^L\left(\frac{\sum_{j} w_{i,j}^{l}}{o_i}\right)^2 \right)
\end{equation}\label{eq:participation_coeff}
and it allows us to estimate whether the total trade of a country is concentrated on a single layer $P =0$ or uniformly distributed across all of them $P =1$~\cite{battiston2014structural}. 

Battiston et al.~\cite{battiston2014structural} propose to use these two metrics to define a space plane with 6 regions (A1-A6), that reflect different ``roles'' that nodes (our countries) can have in a multiplex. In our case, we can interpret these regions as follows:  
\begin{itemize}
    \item \textbf{Multi-product hubs (A6):} main global agents of trade, countries with a large total trade participating in many food categories, $Z(o_{i})\geq2$ and $P_{i}\geq0.66$. 
    \item \textbf{Mixed hubs (A5):} countries with a large total trade, but with moderate participation across food categories, $Z(o_{i})\geq2$ and $0.33<P_{i}<0.66$. 
    \item \textbf{Localized hubs (A4):} countries with a large total trade specialized in a few food categories, $Z(o_{i})\geq2$ and $P_{i}\leq0.33$.
    \item \textbf{Multi-product nodes (A3):} countries with lower trade magnitudes, but participating in many food categories, $Z(o_{i})\leq2$ and $P_{i}\geq 0.66$. 
    \item \textbf{Mixed nodes (A2):} countries with lower trade magnitudes, with a moderate participation across food categories, $Z(o_{i})\leq2$ and $0.33<P_{i}<0.66$. 
    \item \textbf{Localized nodes (A1):} countries with lower trade magnitudes specialized in few food categories,$Z(o_{i})\leq2$ and $P_{i}\le0.33$.
\end{itemize}

\subsection{Simulated production shocks} \label{sec:model}
To estimate the robustness of food trade systems, we quantified the size of the spillover cascades generated by all possible production shortages affecting the system. This implies that, for every year and food category, we quantified the spillover cascades generated after shortages independently starting in each of its global producers. We simulate an initial shock that completely removes ($s = 1$) the production of the affected country ($c_s$) at time $t = 1$ and returns to normal levels afterwards. Additional simulations using partial shocks ($s = 0.3$) are also provided in Supplementary Sec.~2.5.

Given the stochastic nature of the model, we performed \numReps independent repetitions for each single-country shock. Each simulation was run for $T = 50$ time steps, which was sufficient to ensure full absorption of spillover effects in all cases.

\subsection{Shock propagation model}
Our model simulates the immediate aftermath of a production shortage, when countries have not had time to develop new trade partners or to change production, and they only rely on export bans to mitigate supply shortages. As such, the simulation is assumed to operate on a sub-annual scale ($t = 1, 2, \dots, T$), where the structure of $F^y$, demand, and production are constant. 

Under these conditions, countries not facing product shortages will maintain the equilibrium demand defined by \begin{equation}
    d^y_{c_i,f}({t)}= \text{prod}_{c_i,f}(t) + \text{imp}_{c_i,f}(t)-\text{exp}_{c_i,f}(t),
\end{equation}\label{eq:equilibrium}

\noindent
where $\text{prod}_{c_i,f}(t)$, $\text{imp}_{c_i,f}(t)$ and $\text{exp}_{c_i,f}(t)$ indicate respectively the tons of product from food category $f$ produced, imported and exported by country $c_i$ at simulation time $t$~\cite{burkholz2019international}. These can be extracted from the food trade multiplex $F^y$ and production data as follows
\begin{align}
    \text{exp}^y_{c_i, f}(t) &= \sum_{j=1}^{N} W_{c_i,c_j,f}(t), \\
    \text{imp}^y_{c_i, f}(t) &= \sum_{j=1}^{N} W_{c_j,c_i,f}(t), \\
    \text{prod}^y_{c_i, f}(t) &= \text{P}^y_{c_i,f},
\end{align}
where $W^y_{c_i,c_j,f}$ is the weighted adjacency matrix in layer $f$ of the food trade multiplex $F^y$, and $P^y_{c_i,f}$ is the total production of country $c_i$ for food category $f$ in year $y$. Note that $\text{P}^y_{c_i,f}$ is not dependent on $t$ due to our prior assumption of constant production, while the weights of $W^y$ (i.e., tons of product traded) can vary in the time-scale $t$, adapting to shock propagation.

Since our model also assumes that the total demand cannot be modified in the time-scale of the simulation, we can interpret the equilibrium demand in an unperturbed scenario as a constant $\delta^y_{c_i,f}$. This constant can be directly estimated from the trade data as follows 
\begin{equation}
   \delta^y_{c_i,f}= P_{c_i,f}^y + \sum_{j=1}^{N} W_{c_j,c_i,f} -\sum_{j=1}^{N} W_{c_i,c_j,f}.
\end{equation}\label{eq:t=0}

By definition, demand must be positive. Thus, we corrected a few cases where the empirical estimate was $\delta^y_{c_i,f}<0$ by increasing production until $\delta^y_{c_i,f}=0$. This correction only affected \negDemand of the samples evaluated.

After defining the state of the system in an unperturbed scenario ($t=0$), we introduce an exogenous shock in production at $t=1$. This shock consists of reducing the production of an arbitrary country $c_s$ by a certain fraction $s$
\begin{equation}
    \text{prod}^f_{c_s}(t=1) =(1- s)\cdot\text{prod}^y_{c_s, f}(t=0).
\end{equation}
This shock will break the equilibrium demand balance ($d_{c_s,f}(t=1) \neq \delta
_{c_s,f}$), generating a demand deficit in country $c_s$ of $dd_{c_s,f}(t=1)= \delta^y_{c_s,f} - d_{c_s,f} (t=1)= s\cdot\text{prod}^y_{c_s, f}(t=0)$. Countries facing $dd>0$ will always want to minimize their deficit. Thus, they will redirect part (or even all) of their exports into covering domestic consumption. These export restrictions will induce shortages in the countries relying on the removed exports, who will then apply their own restrictions as compensation, further propagating the shock through the trade network. 

Overall, the model can be formalized in the following way: at every timestep $t$ an arbitrary country $c_i$ faces a demand deficit 
\begin{align}
dd_{c_i,f} (t)&= \delta^y_{c_i,f} - d_{c_s,f} (t)= \delta^y_{c_i,f} -\text{prod}_{c_i,f}(t) - \text{imp}_{c_i,f}(t) + exp_{c_i,f}(t)
\end{align}
and it applies the following compensation strategy
\begin{equation}
    \text{exp}_{c_i,f} (t+1)= max\{\exp_{ci,f}(t)-dd_{c_i,f}(t), 0\}. 
\end{equation}

If there are no changes in the production or imports, $dd_{c_i,f}=0$. This implies that $\text{exp}_{c_i,f} (t+1)=\text{exp}_{c_i,f} (t)$, preserving the equilibrium demand balance described before. However, when $dd_{c_i,f}>0$, the compensation strategy will enter into play. Country $c_i$ will apply export bans to its partners until, either completely compensating the demand deficit ($dd_{c_i,f} < \text{exp}_{c_i,f}$), or using all the available exports ($dd_{c_i,f}\geq\text{exp}_{c_i,f}$). Note that when $dd_{c_i,f}\geq\text{exp}_{c_i,f}$, the compensation strategy will not be enough to fully compensate the deficit, and thus the affected country will still have to absorb some residual shortage $dd'_{c_i,f}$. 

Export bans are applied at each timestep following a stochastic propagation rule inspired by Grassia et al.~\cite{grassia2022insights}. In their formulation, Grassia et al. propose a deterministic propagation rule that is inversely proportional to the country's GDP. This is intended to model price competition dynamics in a resource-limited scenario~\cite{grassia2022insights}. We follow their logic by defining the probability of country $c_j$ receiving the consequences of the trade bans applied by $c_i$ as 

\begin{equation}
    p(c_j|c_i) = \frac{GDP_{c_j}^{-1}}{\sum_{c_h \in N_{c_i}} GDP_{c_h}^{-1}},
\end{equation}
where $N_{c_i}$ is the set of importers from $c_i$, i.e., $N_{c_i} = \{\, c_j \;\mid\; W_{c_i,c_j,f}(t) >0 \}$. Then, we choose what trading partners suffer the consequences of the export bans, by randomly sampling without replacement countries $c_j$ until the condition $\sum_{c_{j} } \text{W}_{c_i,c_j,f}(t)\geq dd_{c_i,f}(t)$ is met. Countries with a high $p_{c_i,c_j}$ are more likely to be chosen first, and thus have their supply depleted, while low  $p_{c_i,c_j}$ countries will often be picked last and are more likely to maintain their supplies. However, if $dd_{c_i}>\sum_{c_j \in N_{c_i}}\text{W}_{c_i,c_j,f}$ all $N_{c_i}$ will loose their supplies, and $c_i$  will still face a residual shortage of 
\begin{equation}
    dd'_{c_i,f}(t) =dd_{c_i,f}(t)- \sum_{c_h \in N_{c_i}}\text{W}_{c_i,c_h} (t).
\end{equation}
The residual shortage $dd'_{c_i,f}(t)$ is the main observable of our simulations. 

\subsection{Disruption state}
To contextualize the impact of residual shortages on the food security of countries, we define the ``disruption state'' $\mu_{c_i,c_s,f}$. A country is considered to be disrupted by a spillover cascade starting in $c_s$ if, at some time-point of the simulation, it exceeded a remaining demand deficit $dd'_{c_i,f}$  larger than a predefined tolerance threshold. This can be formalized as 
\begin{equation}
\mu_{c_i,c_s,f} =
\begin{cases}
1 & \text{if } \max\limits_{t \in \mathcal{T}}\;dd'_{c_i,f}(t)>b\cdot\delta^y_{c_i,f}, \\
0 & \text{otherwise,}
\end{cases}\label{eq:failure}
\end{equation}
where $\mathcal{T} =\{t:\{1,\ldots,50\}\}$ is the set of simulated time steps,  and $b\cdot\delta^y_{c_i,f}$ is the tolerance threshold. For simplicity, we assume this threshold to be a fixed fraction $b$ of the country's demand (set to $b=0.01$ in our simulations). The tolerance parameter captures the capacity of a country to buffer temporary supply shortages (e.g., through stocks or other adaptation mechanisms). Thus, this formulation allows us to distinguish residual shortages that are negligible from those that substantially affect the supplies of country $c_i$.

\subsection{Cascade size}\label{meth:casc_size}
We can then use the number of disrupted countries to quantify the size of the spillover cascades triggered by a shock starting in $c_s$. We define it as follows
\begin{equation}
    \text{S}_{c_s} \;=\;100\cdot \frac{\sum\limits_{ c_n \in {\mathcal{N^y}}} \;
\mu_{c_n,c_sf}}{N},
\end{equation}

where $\mathcal{N^y} =\{c_n:\{c_1,...c_N\}\}$ is the set of countries existing in the year $y$ and $N$ is the number of countries in that year. In simpler terms, $S_{c_s}$ reflects the percentage of countries that were disrupted by the shock starting in $c_s$ before it got completely absorbed.

\subsection{Vulnerability}\label{sec:meth_vulnerability}
We define the vulnerability $\beta^f_{X}$ of an arbitrary country $c_i$  as the minimal tolerance threshold that $c_i$ should maintain in order to avoid being disrupted by $X\%$ of the shocks that would otherwise disrupt it. 

From Eq.\ref{eq:failure} we can observe that for country $c_i$ to not be disrupted ($\mu_{c_i,c_s,f}=0$)  by the shock starting in $c_s$, then $ dd'_{c_i,f}(t,c_{s})<b\cdot \delta^y_{c_i,f} \quad \forall t\in\mathcal{T}$, where we change the notation to explicitly state the dependence of the residual shortage on the country starting the shock. With this in mind, the minimum tolerance threshold required for $c_i$ to withstand the shock starting in $c_s$ can be defined as
\begin{equation}
    b^{\text{min}}_{{c_i,f}}(c_s)= \frac{\max\limits_{t \in \mathcal{T}}  dd'_{c_i,f}(t, c_s)}{\delta^y_{c_i,f}}.
\end{equation}

By estimating $b^{\text{min}}_{{c_i,f}}(c_s)$ for all the external shocks disrupting $c_i$, i.e., $S_{c_i}=\{c_s\neq c_i: \mu_{c_i,c_s,f}=1\}$, we can extract a distribution of minimum tolerances for each country. This distribution can be characterized through its cumulative density function $G_{c_i,f} = \text{Pr}[b^{\text{min}}_{c_i,f}<x]$. The CDF describes how likely it is to find a minimal tolerance threshold larger than a specific value $x$. Thus, we can use its inverse to find what would be the $b^{\text{min}}_{c_i,f}$ needed to overcome $X\%$ of the shocks causing a country's failure. This is our vulnerability estimate $\beta_{X}$.

Note that when performing multiple repetitions of the single-country shocks, we estimate the distribution of $b^{\text{min}}_{{c_i,f}}(c_s)$ by considering the average $b^{\text{min}}_{{c_i,f}}(c_s)$ across repetitions with the same $c_s$. 

\bibliography{bibliography.bib}

\section*{Conflict of Interest Statement}
All authors declare to have no conflict of interest. 

\section*{Author Contributions}
Conceptualization: A.F., A.A., R.C. and Y.M.; Methodology: A.F., A.A., R.C. and Y.M.; Investigation: A.F., A.A., R.C. and Y.M.; Writing - original draft: A.F., A.A., R.C. and Y.M.; Writing - review and editing: A.F., A.A., R.C. and Y.M.; Formal analysis and validation: A.F., A.A., R.C. and Y.M.; Software: A.F. All authors have read, edited, and approved the final manuscript.

\section*{Funding}
A.F., A.A., and Y.M were partially supported by the Government of Aragón, Spain, and ``ERDF A way of making Europe'' through grant E36-23R (FENOL), and by Ministerio de Ciencia, Innovación y Universidades, Agencia Española de Investigación (MICIU/AEI/ 10.13039/501100011033) Grant No. PID2023-149409NB-I00. A.A. acknowledges support from the grant RYC2021-033226-I funded by MICIU/AEI/10.13039/501100011033 and the European Union NextGenerationEU/PRTR.

\section*{Data Availability}
All data used in this study are publicly available from the following sources. The normalized annual bilateral trade flow matrices, production records, and GDP data were all extracted from FAOSTAT~\cite{FAO_trade,FAO_gdp}. The hierarchical food classification was extracted from the FoodEx2 database~\cite{fao1993codex}.

\section*{Code Availability}
The code is available in this repository: \url{https://github.com/AFosch/food_trade_shocks}.

\end{document}